\documentclass[conference]{IEEEtran}
\IEEEoverridecommandlockouts
\usepackage{cite}
\usepackage{amsmath,amssymb,amsfonts}
\usepackage{algorithmic}
\usepackage{graphicx}
\usepackage{textcomp}
\usepackage{xcolor}
\usepackage[T1]{fontenc} 
\def\BibTeX{{\rm B\kern-.05em{\sc i\kern-.025em b}\kern-.08em
    T\kern-.1667em\lower.7ex\hbox{E}\kern-.125emX}}
\usepackage{fancyhdr}

\begin{document}

\title{Best CNTFET Ternary Adders? }

\author{\IEEEauthorblockN{ Daniel Etiemble}
\IEEEauthorblockA{\textit{Computer Science Laboratory (LRI)} \\
\textit{Paris Saclay University}\\
Saint Aubin, France \\
de@lri.fr}

}
\maketitle

\begin{abstract}
The MUX implementation of ternary half adders and full adders using predecessor and successor functions lead to the most efficient efficient implementation using the smallest transistor count. These designs are compared with the binary implementation of the corresponding half adders and full adders using the MUX technique or the typical complementary CMOS circuit style. The transistor count ratio between ternary and binary implementations is always greater than the information ratio ($log_2(3)/log_2(2)$ = 1.585) between ternary and binary wires. 

\end{abstract}


\section{Introduction}\label{sec1}
Many ternary half adders and full adders have been presented in the last decade. The most significant papers are \cite{Lin}, \cite{Samadi}, \cite{Sahoo}, \cite{Jaber}. \cite{Mirzaee}, \cite{Ebrahimi}, \cite{Kesh}.
In \cite{eti1}, we have compared different implementations of quaternary adders. The best quaternary one has been presented in \cite{Roosta}. It is based on the use of multiplexers. In this paper, we show that this approach also leads to the best implementation of ternary half and full adders. The design is based on CNTFET technology. This technology is far from being a mature technology. As of 2020, FinFET technology integrates millions of times more transistors than CNTFET technology. However, CNTFET has a big advantage for designing multivalued circuits. While it basically uses the typical CMOS circuit styles, the threshold levels of the different multivalued gates can be got by adjusting the diameter of each used transistor. This technology is used in this paper.
We first present the MUX based design of the ternary half adder. Then we present the full adder design. These implementation are compared with the similar MUX based implementation and the conventional implementation of the binary versions.

\section{Ternary Half Adders}
  Table \ref{T1} presents the truth table of the ternary half-adder.

Ternary half adders and full adders have ternary inputs and outputs and binary carry inputs and outputs. 
\begin{itemize}
\item Ternary values (0,1,2) corresponding to 0, $V_{dd}/2$ and $ V_{dd}$ voltage levels
\item According to the circuit style that is used, the binary values  may have levels 0 and $ V_{dd}/2$ or 0 and $V_{dd}$. 
\end{itemize}

The technique used in \cite{Roosta} is based on multiplexers. In this approach, the carry signals are used as control inputs of MUX: they are never used as input values of these MUX. The binary levels are thus 0 and $V_{dd}$. The corresponding  binary values are 0 and 2.

\begin{table}
\centering
\caption{Half adder truth table}
\begin{tabular}{|c|c|c|c|c|c|c|c|c|}
  \hline
 & \multicolumn{3}{|c|}{SUM}&&\multicolumn{3}{|c|}{CARRY}\\
  \hline
X/Y&0&1&2&&0&1&2\\
  \hline
0&0&1&2&&0&0&0\\
1&1&2&0&&0&0&1\\
2&2&0&1&&0&1&1\\
  \hline
\end{tabular}
\label {T1}
\end{table}

Computation of the Half Adder SUM is processed by the following rules:
\begin{itemize}

\item When X = 0, SUM = Y
\item When X = 1, SUM = (Y+1) mod (3). The corresponding circuit is called successor circuit.
\item When X = 2, SUM = (Y-1) mod (3). The corresponding circuit is called predecessor circuit.
\item According to X, a ternary MUX provides the correct SUM output.
\end{itemize}

\begin{figure}[htbp]
\centerline{\includegraphics  [width =8 cm]{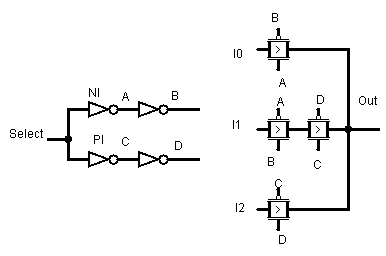}}
\caption{Threshold detectors and ternary MUX (Ternary inputs and ternary control)}
\label{TMUX}
\end{figure}

Threshold detectors circuits are implemented by the inverters NI and PI, according to Table 2. These inverters are generally called NTI and PTI, but this is confusing as they are binary inverters: the outputs are binary ones and they only have one threshold level. The only difference with typical binary inverters is the specific threshold levels. Assuming$ V_{dd}$, $V_{dd}/2$ and 0 ternary levels, the threshold levels are $V_{dd}/4$ and $3V_{dd}/4$ when the threshold level of a binary inverter is $ V_{dd}/2$. The NI and PI threshold levels are obtained by choosing the appropriate diameter for the different CNTFET transistors. They are shown in left part of Figure \ref{TMUX}. The implementation of the ternary MUX is shown in right part of Figure \ref{TMUX}. 

\begin{table}
\centering
\caption{NI and PI binary functions}
\begin{tabular}{|c|c|c|c|c|c|c|c|c|}
  \hline
 &NI&PI\\
  \hline
0&2&2\\
1&0&2\\
2&0&0\\
  \hline
\end{tabular}
\label {T2}
\end{table}

\begin{figure}[htbp]
\centerline{\includegraphics  [width =9 cm]{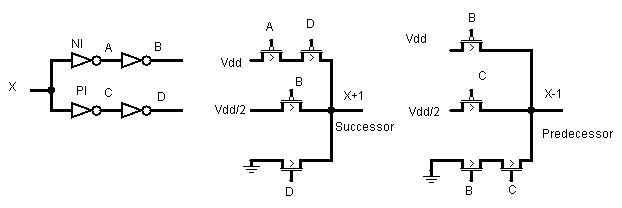}}
\caption{Successor and Predecessor Circuits (2 power supplies)}
\label{PreSuc2PS}
\end{figure}

\begin{figure}[htbp]
\centerline{\includegraphics  [width =9 cm]{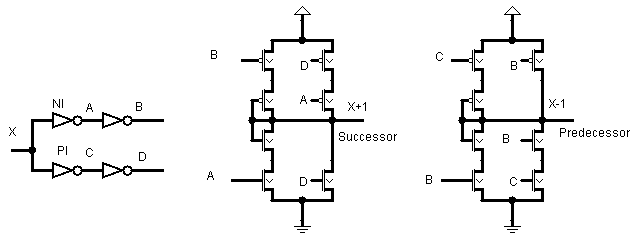}}
\caption{Successor and Predecessor Circuits (1 power supply)}
\label{PreSuc1PS}
\end{figure}

Figure \ref{PreSuc2PS} presents the Successor and Predecessor circuits with two power supplies ($V_{dd}$ and $V_{dd}/2)$. A, B, C, D outputs of NI and PI inverters control the different transistors: for each ternary input X, only one path is active between $V_{dd}$ or $V_{dd}/2$ or ground and the corresponding output. The drawback of this approach is the supplementary power supply.
Figure \ref{PreSuc1PS} presents the successor and predecessor circuits with only one power supply $(V_{dd})$. The intermediate level is got through a voltage divider by using resistor-like transistors. The drawback of this approach is a static power dissipation when the output is 1.

\begin{figure}[htbp]
\centerline{\includegraphics  [width =6 cm]{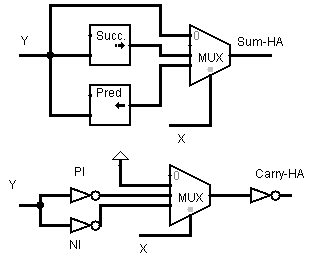}}
\caption{Ternary Half Adder }
\label{TernaryHA}
\end{figure}

Figure \ref{TernaryHA} presents the ternary half-adder circuit. The carry output is:
\begin{itemize}
\item	When X = 0, $C_{out}$ = 0
\item	When X = 1, $C_{out}$ = 1 if Y = 2 ($\overline{PI(Y)} = 2$) according to Table 2)
\item	When X = 2, $C_{out}$ = 1 if  $Y \geq 1$ ($\overline{NI(Y)} = 2$) according to Table 2)
\end{itemize}

Table \ref{T3} presents the transistor count for the half adders.

\begin{table}
\centering
\caption{Transistor count for the Ternary Half Adder}
\begin{tabular}{|c|c|c|c|c|c|c|c|c|}
  \hline
&2 PS&1 PS\\
  \hline
 Succ-Pred&8&14\\
MUX&16&16\\
PI-NI&16&16\\
NOT&2&2\\
  \hline
TOTAL&42&48\\
  \hline
\end{tabular}
\label {T3}
\end{table}

Table \ref{T4} presents the transistor count for different ternary half adders proposed in the last ten years. 

\begin{table}
\centering
\caption{Transistor count of different HAs}
\begin{tabular}{|c|c|c|c|c|c|c|c|c|}
  \hline
 Ternary Half Adder &\cite{Lin}& \cite{Samadi} &\cite{Sahoo} &\cite{Jaber}&New 2 PS&New 1PS\\
  \hline\
Transistor count&136&112&112&85&42&48\\
  \hline
\end{tabular}
\label {T4}
\end{table}

\section{Ternary Full Adders}
When $C_{in}$ = 0, the full adder truth table was presented in Table 1 (half-adder).When $C_{in}$ = 1, the truth table is given in Table \ref{T5}.

\begin{table}
\centering
\caption{Full adder truth table when $C_{in} = 1$}
\begin{tabular}{|c|c|c|c|c|c|c|c|c|}
  \hline
 & \multicolumn{3}{|c|}{SUM}&&\multicolumn{3}{|c|}{CARRY}\\

  \hline\
X/Y&0&1&2&&0&1&2\\
  \hline\
0&1&2&0&&0&0&1\\
1&2&0&1&&0&1&1\\
2&0&1&2&&1&1&1\\
  \hline
\end{tabular}
\label {T5}
\end{table}

Still using the MUX approach, the SUM output of the full adder is given by
\begin{itemize}
\item	If $C_{in}$ = 0, then $SUM_{FA} = SUM_{HA}$ else $SUM_{FA} = (SUM_{HA} +1) mod(3)$
\end{itemize}
Another approach directly computes $SUM_{FA}$ as a function of $C_{in}$ and X.
\begin{itemize}

\item	When X = 0: if $C_{in}$ = 0 then SUM = Y else SUM = (Y+1) mod (3).
\item	When X = 1, if $C_{in}$ = 0 then SUM = (Y+1) mod (3) else SUM = (Y-1) mod (3).
\item	When X = 2, if $C_{in}$ = 0 then SUM= (Y-1) mod (3) else SUM =Y. 
\end{itemize}
$C_{out1}$  is the carry output when $C_{in}$ = 1:
\begin{itemize}
\item	When X = 0, $C_{out1}$ = 1 if Y = 2 ($\overline{PI(Y)} =2$  according to Table 2)
\item	When X = 1, $C_{out1}$ = 1 if $Y \geq 1$ ($\overline{NI(Y)} =2$ according to Table 2)
\item	When X = 2, $C_{out1}$ = 1
\end{itemize}
If $C_{in}$ = 0 then $C_{outFA}$  = $C_{outHA}$  else $C_{outFA}$  = $C_{out1}$

\begin{figure}[htbp]
\centerline{\includegraphics  [width =7 cm]{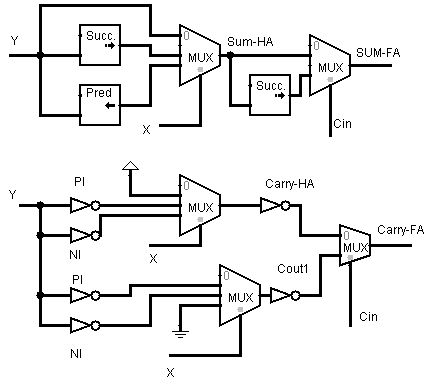}}
\caption{Ternary Full Adder(version 1)}
\label{TernaryFA}
\end{figure}

\begin{figure}[htbp]
\centerline{\includegraphics  [width =7  cm]{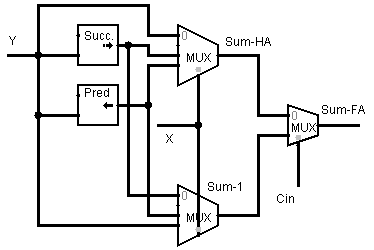}}
\caption{Ternary Full Adder (version 2)}
\label{TernaryFA2}
\end{figure}

\begin{figure}[htbp]
\centerline{\includegraphics  [width =4 cm]{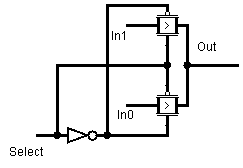}}
\caption{Typical MUX2 with binary control)}
\label{MUX2B}
\end{figure}

Figure \ref{TernaryFA} and Figure \ref{TernaryFA2} use two new different types of MUXes:
\begin{itemize}
\item MUXes with ternary inputs and binary control;
\item MUXes with binary inputs and binary control.
\end{itemize}
Both types use the same typical MUX2 binary circuit (Figure \ref{MUX2B})

Table \ref{T6} presents the transistor counts for the two versions of the ternary full adder with two power supplies. 
Table \ref{T7} provides the same information for one power supply.

\begin{table}
\centering
\caption{Transistor counts for thee ternary full adder (2 power supplies)}
\begin{tabular}{|c|c|c|c|c|c|c|c|c|c|c|}
 
  \hline\
&Carry&SUM V1 	&SUM V2	&FA-V1	&FA-V2	\\
  \hline\
SUCC- PRED	&0&	12&	8		&	&	\\
MUX3	&16	&8&	16			&	&\\	
MUX2 &4 &4&4&&\\
PI-NI&	0&	24	&16	&		&		\\
NOT	&6&	2&	2&	&				\\
  \hline
TOTAL&	26&	50&	46	&76	&72	\\
\hline
\end{tabular}
\label {T6}
\end{table}

\begin{table}
\centering
\caption{Transistor counts for the ternary full adder (1 power supply)}
\begin{tabular}{|c|c|c|c|c|c|c|c|c|c|c|}

  \hline\
&Carry&SUM V1	&SUM V2	&FA-V1	&FA-V2\\
  \hline\
SUCC- PRED	&0&	21&14&		&	\\
MUX3	&16	&8&16&&\\
MUX2 &4&4&4&&\\
PI-NI&	0&			24	&16&&		\\
NOT	&6	&		2&	2&	&	\\
  \hline
TOTAL&	26	&59	&52&	83&	78\\
\hline
\end{tabular}
\label {T7}
\end{table}

Table \ref{T8} presents the transistor count for different ternary full adders proposed in the last decade.

\begin{table}
\centering
\caption{Transistor count of different ternary FAs}
\begin{tabular}{|c|c|c|c|c|c|c|c|c|}
  \hline
 Ternary Full Adder &\cite{Mirzaee}& \cite{Ebrahimi} &\cite{Kesh} &Proposed\\
  \hline\
TC (2 power supplies)&&106&132&72\\
TC (1 power supply)&142&&&78\\
  \hline
\end{tabular}
\label {T8}
\end{table}

\section{Comparing with binary adders}
\subsection{Binary Half Adder}
\begin{figure}[htbp]
\centerline{\includegraphics  [width =4 cm]{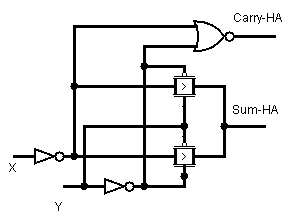}}
\caption{Binary Half Adder (MUX technique)}
\label{BHAMUX}
\end{figure}

\begin{figure}[htbp]
\centerline{\includegraphics  [width =7 cm]{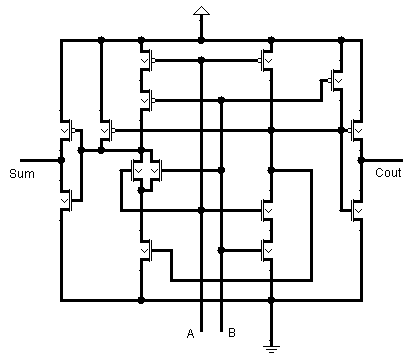}}
\caption{14 T binary Half Adder)}
\label{14TBHA}
\end{figure}

Using the same MUX technique, the binary half-adder is shown in Figure \ref{BHAMUX}. It uses 12 T. Permuting the MUX inputs and adding an output inverter would lead to 14 T with restored output levels. The transistor count is then the same as a typical conventional approach used in standard cell libraries \cite{VLSI}, as shown in Figure \ref{14TBHA} . 

\subsection{Binary Full Adder}
\begin{figure}[htbp]
\centerline{\includegraphics  [width =6  cm]{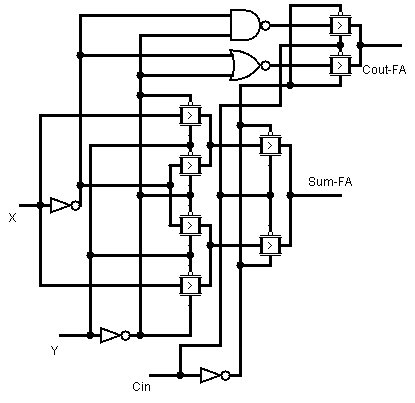}}
\caption{Binary Full Adder (MUX technique)}
\label{BFAMUX}
\end{figure}

\begin{figure}[htbp]
\centerline{\includegraphics  [width =7  cm]{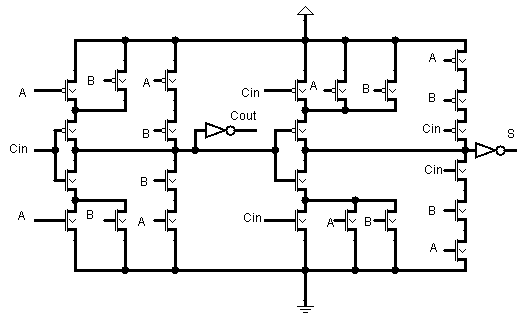}}
\caption{28 T binary Full Adder)}
\label{28TBFA}
\end{figure}

With the MUX approach, the binary full adder is presented in Figure \ref{BFAMUX}. It corresponds to the following rules:
\begin{itemize}
\item If $C_{in}$ = 0 then $SUM_{FA}$ = X XOR Y else X NXOR Y. XOR and NXOR functions are implemented using MUXes.
\item If $C_{in}$ = 0 then $C_{out}$ = X.Y else $C_{out}$ = X+Y.
\end{itemize}
The MUX based full adder has 30 T. A version with restored output levels would have 34 T. It could be outlined that the MUX approach has more transistors than the typical conventional full adder with complementary CMOS shown in Figure \ref{28TBFA}. Many proposals using less than 28 T have been proposed in the literature. 

\section{Conclusion}
The proposed ternary half adders and full adders have less transistors than all the previously proposed ones. It seems that the MUX approach with successor and predecessor circuits is the best one to implement ternary arithmetic circuits. It looks like the transistor counts are close to the minimal possible value.
The only valid comparison between ternary and binary circuit is based on the information ratio. According to Shannon theory of information, when N events have the same probability to occur, the corresponding amount of information is $I = log_2 (N)$ bits (or Shannon). When N = 2, I = 1 bit. When N = 3, I = 1.585 bits. A ternary wire carries 1.585 times the amount of information of a binary one. This 1.585 information ratio must be used to compare binary and ternary circuits. For instance, an 8-bit binary adder can be compared to a 5-trit ternary adder as they process approximately the same amount of information. 8/5 is close to 1.585. The difference results from rounding issues.
Considering the most conservative implementation of binary circuits, the transistor count ratio between ternary and binary half adder is 42/14 = 3 and 72/28 = 2.57 for the full adder when using two power supplies for the ternary case. With only one power supply, the ratios are respectively 48/14 = 3.4 and 78/28 =  2.8. Both ratios are greater than 1.585. It means that the best ternary implementation leads to more transistors, more chip area, more interconnects and more power dissipation than the corresponding conservative binary ones. 
These results are not surprising. 3 is not the best base for computation \cite{eti2} and multivalued circuits are restricted to a small niche \cite{eti3}.

\end{document}